\documentstyle[12pt]{article}

\def\thefootnote{\fnsymbol{footnote}}
\renewcommand{\thefootnote}{\alph{footnote}}

\newcommand{\rref}[1]{(\ref{#1})}
\newcommand{\beqn}{\begin{equation}}
\newcommand{\eeqn}{\end{equation}}
\newcommand{\beqarr}{\begin{eqnarray}}
\newcommand{\eeqarr}{\end{eqnarray}}

\newcommand{\matc}{\begin{array}{c}}
\newcommand{\matcc}{\begin{array}{cc}}
\newcommand{\matccc}{\begin{array}{ccc}}
\newcommand{\matcccc}{\begin{array}{cccc}}
\newcommand{\emat}{\end{array}}
\newcommand{\df}{\stackrel{\rm def}{=}}
\newcommand{\SOS}{SO(3,3\,|{\bf Z})}
\newcommand{\bn}{\tilde{n}}
\newcommand{\bm}{\tilde{m}}
\newcommand{\bM}{\widetilde{M}}

\begin{document}

\begin{titlepage}

October 1998         \hfill
\begin{center}
\hfill    UCB-PTH-98/48 \\
\hfill    LBNL-42383	 \\

\vskip .15in
\renewcommand{\thefootnote}{\fnsymbol{footnote}}
{\large \bf Dualities of the Matrix Model \\
from T-Duality of the Type II String}\footnote{This 
work was supported in
part by the Director, Office of Energy Research, Office of High Energy and 
Nuclear Physics, Division of High Energy Physics of the U.S. Department of 
Energy under Contract DE-AC03-76SF00098 and in part by the National Science
Foundation under grant PHY-95-14797}
\vskip .15in

Daniel Brace\footnote{email address: brace@thwk2.lbl.gov}, 
Bogdan Morariu\footnote{email address: morariu@thsrv.lbl.gov} and 
Bruno Zumino\footnote{email address: zumino@thsrv.lbl.gov}
\vskip .15in

{\em 	Department of Physics  \\
	University of California   \\
				and	\\
	Theoretical Physics Group   \\
	Lawrence Berkeley National Laboratory  \\
	University of California   \\
	Berkeley, California 94720}
\end{center}
\vskip .15in

\begin{abstract}
We investigate in the Matrix theory framework, the subgroup of
dualities of the DLCQ of M-theory compactified on 
three-tori, which  corresponds to T-duality in the auxiliary 
\mbox{Type II} string theory. We show how these dualities  are 
realized in the supersymmetric 
Yang-Mills gauge theories on dual noncommutative three-tori.
\end{abstract}
\end{titlepage}
%THIS PAGE (PAGE ii) CONTAINS THE LBL DISCLAIMER
%TEXT SHOULD BEGIN ON NEXT PAGE (PAGE 1)
\renewcommand{\thepage}{\roman{page}}
\setcounter{page}{2}
\mbox{ }

\vskip 1in

\begin{center}
{\bf Disclaimer}
\end{center}

\vskip .2in

\begin{scriptsize}
\begin{quotation}
This document was prepared as an account of work sponsored by the United
States Government. While this document is believed to contain correct
 information, neither the United States Government nor any agency
thereof, nor The Regents of the University of California, nor any of their
employees, makes any warranty, express or implied, or assumes any legal
liability or responsibility for the accuracy, completeness, or usefulness
of any information, apparatus, product, or process disclosed, or represents
that its use would not infringe privately owned rights.  Reference herein
to any specific commercial products process, or service by its trade name,
trademark, manufacturer, or otherwise, does not necessarily constitute or
imply its endorsement, recommendation, or favoring by the United States
Government or any agency thereof, or The Regents of the University of
California.  The views and opinions of authors expressed herein do not
necessarily state or reflect those of the United States Government or any
agency thereof, or The Regents of the University of California.
\end{quotation}
\end{scriptsize}

\vskip 2in

\begin{center}
\begin{small}
{\it Lawrence Berkeley National Laboratory is an equal opportunity employer.}
\end{small}
\end{center}

\newpage
\renewcommand{\thepage}{\arabic{page}}

\setcounter{page}{1}
\setcounter{footnote}{0}

% main text is here

\section{Introduction}
\label{Intro}

It was conjectured by Susskind~\cite{LS} that each momentum sector of the
discrete light cone quantization (DLCQ) of M-theory is described 
by a maximally supersymmetric
Matrix model~\cite{CH,FL,BRR} with the momentum identified 
with the rank of the gauge 
group. The conjecture was further clarified by Sen and 
Seiberg~\cite{AS,NS}. They used an infinite boost and a 
compensating rescaling to show that
the DLCQ Hamiltonian of the original
M-theory is given by the Hamiltonian of an auxiliary M-theory compactified 
on a vanishingly small space-like circle of radius $R$. 
This is then equivalent to a
weakly coupled Type IIA string theory,  which will be referred to,
following Sen~\cite{AS4},
as the auxiliary Type II string theory. At the same time,
the original light-cone 
momentum is mapped into Ramond-Ramond D0 brane charge.
The string coupling and string mass scale are given by the
$R \rightarrow 0$ limit  of
\[
g_S= M_P^{3/2}(LR)^{3/4},~m_S=M_P^{3/2}L^{3/4}R^{-1/4}.
\]
In this limit as proposed by Witten~\cite{W}, 
and discussed extensively in~\cite{DKPS},
the dynamics of $n$ D0 branes
is determined by the maximally supersymmetric Matrix 
model~\cite{CH,FL,BRR}.

Toroidal compactification of M-theory can be obtained by considering 
Matrix theory on the covering space of the torus 
and imposing a periodicity constraint on
the dynamical variables~\cite{BFSS,WTc,GRT}. The constrained system
is formally equivalent to a $U(n)$ super Yang-Mills (SYM) gauge theory
on a  dual torus. On the other hand upon compactification on a 
$d$-dimensional torus $T^d$ M-theory has additional moduli
from the three form of eleven dimensional supergravity.
Connes, Douglas and Schwarz~\cite{CDS}, conjectured that these
moduli correspond to the deformation parameters $\Theta_{ij}$ 
of a noncommutative super Yang-Mills (NCSYM) gauge theory.
Further studies of this subject followed 
in~\cite{DH,PWW,PW,C,CK,KO,AS1,AAS,PMH,DBsym,AS2,LLS,BB,AAS2}.

In~\cite{CDS}, where compactification on a two-torus was considered
in some detail,
it was suggested that the  $SL(2,{\bf Z})$ noncommutative 
duality group of the NCSYM gauge theory~\cite{R1,R2,R3,R4,R5} corresponds to 
the T-duality in the DLCQ direction and one of the space-like 
compact directions of M-theory. Later Rieffel and
Schwarz~\cite{AS1} showed that NCSYM gauge theories 
on higher dimensional tori have an $SO(d,d\,|{\bf Z})$
duality, and conjectured that this is the realization,  in the NCSYM theory, 
of the auxiliary Type II string theory T-duality.

In this paper we investigate this conjecture 
for compactifications on a three-torus.
First we extend the method used in~\cite{PMH,BB} to construct
twisted bundles to the three-torus.
Then we show  explicitly how to construct an action of the duality 
group $SO(3,3\,|{\bf Z})$ on NCSYM theories. Under these duality 
transformations the rank of the gauge group and the magnetic flux numbers 
transform together in a Weyl spinor representation, and the deformation 
parameters transform by fractional transformations. We also obtain 
the transformation properties, under the duality group, of the 
gauge coupling and the metric. We can then directly compare these relations 
with the string theory T-duality predictions.

In the next section  
we review the standard toroidal Matrix compactification
leading to a SYM gauge theory on the dual torus. Then we present the 
conjecture~\cite{CDS}, that in the presence of  
nonvanishing NS antisymmetric moduli $B_{ij}$, the translation generators 
implementing the quotient condition do not
commute, such that Matrix compactification leads
to a noncommutative super Yang-Mills  gauge theory on 
a dual noncommutative torus.

In Section~\ref{QB}, we study adjoint quantum bundles on noncommutative 
tori of arbitrary dimension which admit a constant 
curvature which is not valued in the $su(n)$ subalgebra 
and have transition functions of a special simple form.

In Section~\ref{Adj} we show how to expand the sections of  the adjoint 
bundle of a $U(n)$ gauge theory in terms of matrix valued functions
on a dual noncommutative torus.  The dual deformation parameter $\Theta'$
lies on the same $SO(d,d\,|{\bf Z})$ orbit as the original~$\Theta$.
We perform most of the calculations on tori of arbitrary dimension,
but later in the paper we concentrate on the two and three-tori.

In  Section~\ref{T2}, we describe the quantum bundles corresponding to the
two dimensional compactification.  We rewrite some of the known two
dimensional relations in a form that admits immediate generalization to 
higher dimensions.  
Section~\ref{T3} contains the solution for arbitrary bundles over 
three dimensional tori.  

In Section~\ref{NCSYM}, we consider the  noncommutative SYM action 
on a twisted quantum bundle after a brief description of 
the quantum integral. 

In Section~\ref{DSYM}, we show how to manipulate the NCSYM action of a $U(n)$
theory on a twisted three-torus such that it is formally equivalent to
a $U(q)$ action on a trivial bundle over  a dual torus,
where $q$ is the greatest 
common divisor of $n$ and the magnetic
fluxes of the bundle.
More generally, two NCSYM theories are equivalent if their
rank and magnetic fluxes and deformation parameters are on the same 
orbit of $SO(3,3\,|{\bf Z})$. The rank and the magnetic fluxes transform
in an integral Weyl spinor representation of the group.
Related results were obtained in~\cite{AS2} using a more 
abstract mathematical language.
In the last section we show that this 
duality is the low energy
remnant of the T-duality of Type II string theory.

Finally in the appendix we prove a theorem showing that the chiral spinor 
representations of $SO(d,d\,|{\bf Z})$ are integral, and also show that 
the spinor representation of $SO(3,3\,|{\bf Z})$ is in fact $SL(4,{\bf Z})$.

\section{Matrix Compactification}
\label{MC}

In this section we present a review of Matrix theory compactification.
In the limit of large string mass the dynamics of $n$ D0 branes, 
in uncompactified space-time,
is determined by the maximally supersymmetric Matrix 
action~\cite{CH,FL,BRR},
\[
{\cal S}^{D0} = \frac{1}{2 g_{S}} \int dt\,{\rm tr} (
 \sum_M \dot{X}^M  \dot{X}^M+ \frac{1}{(2\pi)^2}
\sum_{M < N} [X^M,X^N] [X^M,X^N]+{\rm fermions} ).
\]
This action is obtained by dimensional reduction 
of the ten dimensional \mbox{${\cal N}=1$} SYM gauge theory.
Alternatively we could work with the IKKT functional~\cite{IKKT}
obtained by dimensionally reducing, in all directions 
including time,
the  Euclidean ten dimensional SYM action.

The compactification of Matrix theory 
on a $d$-dimensional torus is obtained by considering an
infinite number of D0 branes living on ${\bf R}^d$, the covering space 
of the torus, and then
imposing the following quotient conditions~\cite{BFSS,WTc}.
\[
{\cal U}_i^{-1} X^I {\cal U}_i = 2\pi e^I_i + X^I,~i,I= 1,\ldots,d,
\]
\[
{\cal U}_i^{-1} X^a {\cal U}_i = X^a,~a= d+1,\ldots,9.
\]
Here $I$ runs over the compact directions, and the $e^I_i$ form
a basis defining the compactification lattice. The ${\cal U}_i$'s are unitary
operators.
One can define new matrix coordinates
\[
X^i = e^i_I X^I,
\]
which obey the simpler quotient conditions
\beqn
{\cal U}_i^{-1} X^j {\cal U}_i = 2\pi \delta_i^j + X^j. \label{quotient}
\eeqn
In terms of the new variables the action takes the form
\[
{\cal S}^{D0} = \frac{1}{2 g_{S}} \int dt~{\rm Tr} \left(
G_{ij} \dot{X}^i  \dot{X}^j+ 
\frac{1}{2}\frac{1}{(2\pi)^2} G_{ij} G_{kl} [X^i,X^k] [X^j,X^l]+  \right.
\]
\beqn
 \sum_{a} \dot{X}^a  \dot{X}^a+
 \frac{1}{(2\pi)^2}\sum_{a} G_{ij} [X^i,X^a] [X^j,X^a]+   \label{D0action}
\eeqn
\[
\left.
 \frac{1}{(2\pi)^2}\sum_{a<b} [X^a,X^b] [X^a,X^b]
+{\rm fermions}\right),
\]
where we have introduced the metric $G_{ij}=\sum_I e_i^I  e_j^I$. 
In~(\ref{D0action}), the trace
over infinite dimensional matrices is formally divided
by the infinite order the quotient group ${\bf Z}^d$.

The original solution of the quotient condition assumed that the 
translation operators commute
\[
[{\cal U}_i,{\cal U}_j]=0.
\]
The standard way to solve~\rref{quotient} is to introduce an
auxiliary Hilbert space on which $X^i$'s and ${\cal U}_i$'s act. In the 
simplest case this is taken to be  the space of 
functions on a $d$-dimensional torus taking values in ${\bf C}^n$.  
Then if we let the ${\cal U}_i$'s be the generators of the algebra of 
functions on the torus
\[
{\cal U}_i=e^{i\sigma_i},
\]
where $\sigma_i$ are coordinates on the covering space of the torus,
the $X^i$'s satisfying~(\ref{quotient}) must be covariant derivatives
\beqn
X^j = -2\pi i \, D^j = -2\pi i (\partial^j -i A^j({\cal U}_k)).
\label{connection}
\eeqn
The partial derivative is with respect to $\sigma_j$, and $A^j$
are $n$-dimensional hermitian matrices. The action~(\ref{D0action}) 
can be rewritten as a $d$-dimensional SYM action, by replacing
the $X^i$'s with covariant derivatives as above, and rewriting the
trace over the infinite dimensional matrices 
as
\[
 {\rm Tr} = \int \frac{d^d\sigma}{(2\pi)^{d}}\
{\rm tr}.
\]
Here tr is an $n$-dimensional trace, and the new coordinates $\sigma_i$ are
to be integrated from zero to $2\pi$. The action becomes
\[
{\cal S}^{D0} = \frac{{(2\pi)}^{2-d}}{4 g_{S} \sqrt{\det(G^{ij})} }
 \int dt
\int  {d^d\sigma}\, \sqrt{\det( G^{ij})}
~{\rm tr} \left(
G_{\mu \nu} G_{\xi \rho} [D^{\mu},D^{\xi}] 
[D^{\nu},D^{\rho}]- \right.
\]
\[
 \sum_{a} G_{\mu \nu} [D^{\mu},X^a] [D^{\nu},X^a]+
 \sum_{a<b} [X^a,X^b] \left. [X^a,X^b]
+{\rm fermions}      \right),
\]
where the scalar fields $X^a$ have been rescaled by a factor of $2\pi$.
We have written the action in standard form\footnote{Note that the 
positions of all the indices are  switched. For example the metric has
upper indices. This just reflects that we have performed a T-duality
under which the metric is replaced with the inverse metric. Another way to 
understand the index position is that T-duality is a canonical transformation
which exchanges coordinates and momenta and therefore reverses the 
index structure.} so that we can read  off 
the SYM gauge coupling 
\beqn
g_{SYM}^2= g_{S} ~ (2\pi)^{d-2} \, \sqrt{\det(G^{ij})}.  \label{ggg}
\eeqn
Thus  the 
gauge coupling $g_{SYM}^2$ equals the string coupling on the T-dual torus. 
The square root factor accounts for the expected dilaton shift under 
T-duality.
%Note that compactification of the IKKT model gives the 
%finite temperature action of the Matrix model.

Following~\cite{CDS} we consider the general 
case when the unitary operators ${\cal U}_i$ 
do not commute. Consistency of the quotient conditions requires
that the ${\cal U}_i$'s must commute up to a phase
\[
{\cal U}_i {\cal U}_j= e^{-2\pi i \Theta_{ij}} {\cal U}_j {\cal U}_i.
\label{calUUcom}
\]
Connes, Douglas and Schwarz conjectured that the deformation 
parameters $\Theta$ correspond
to certain moduli of the compactification of the DLCQ of M theory 
on tori.
If $\gamma^{ij-}$ represents a three cycle wrapped around the 
transversal directions $x^{i}$ and $x^{j}$ and the  light cone 
direction $x^{-}$, then 
\[
\Theta_{ij} = \frac{1}{(2\pi)^3} \int_{\gamma^{ij-}}C,
\]
where $C$ is the antisymmetric three 
form of eleven dimensional supergravity. 
Written in terms of the auxiliary type IIA string theory variables,
\[
\Theta_{ij} = \frac{1}{(2\pi)^2} \int_{\gamma^{ij}}B,
\]
where $B$ is the NS two form.
In the noncommutative case it is convenient to introduce another set of
translation operators $U_i$ which satisfy
\beqn
U_i U_j= e^{2\pi i \Theta_{ij}}  \label{UU}
  U_j  U_i.
\eeqn
The $U_i$'s generate the algebra of functions on a quantum torus. 
We will denote this algebra, ${\cal A}_{\Theta}$. 
For an expanded discussion of this and other issues in noncommutative
geometry see~\cite{CON,DB}.
This algebra can 
be realized as
a subalgebra of the quantum plane algebra, which is generated by
$\sigma_i$ satisfying
\beqn
[\sigma_i,\sigma_j]=-2\pi i \Theta_{ij}.  \label{sigcom}
\eeqn
Then we realize the generators of  ${\cal A}_{\Theta}$ as
\[
U_i \df e^{i \sigma_i}.
\]
To realize the ${\cal U}_i$ generators we also introduce 
partial derivatives satisfying\footnote{Just
as in the classical case, one can also introduce quantum 
exterior forms $d\sigma_i$, which anti-commute with each other and commute
with all other variables.}
\[
[\partial^i,\sigma_j]=\delta^{i}_{j},~[\partial^i,\partial^j]=0.
\]
Now we can write the ${\cal U}_i$ generators as
\[
{\cal U}_i=e^{i \sigma_i-2 \pi \Theta_{ij}\partial^j}.
\]
Note that both $\sigma_i$ and $\partial^i$ act as translation generators
on the $\sigma_i$'s, and the exponent in the ${\cal U}_i$'s is just the linear 
combination that commutes with all the $\sigma_i$'s.  Thus
\[
[{\cal  U}_i, U_j]=0.
\]
For vanishing $\Theta$ we see that $U_i$ and ${\cal U}_i$ coincide.

The simplest example of solutions of the quotient conditions~(\ref{quotient})
are quantum connections on trivial bundles
\beqn
X^j = -2\pi i \, D^j = -2\pi i (\partial^j -i A^j(U_k)).
\label{NCconnection}
\eeqn
In the noncommutative case the matrix elements of $A^j$ are elements of
${\cal A}_{\Theta}$. Again using the representation~(\ref{NCconnection})
of $X^i$ in the Matrix model action we obtain a NCSYM 
action~\cite{CR}. However we will postpone writing this
action until we study more general solutions which are
connections on nontrivial bundles.

\section{Twisted Quantum Bundles on Tori}
\label{QB}

In this section we construct quantum $U(n)$ bundles on 
$d$-dimensional noncommutative tori
which admit 
constant curvature connections with vanishing $su(n)$ curvature.
This is done by finding explicit 
transition functions compatible  with such a connection. We employ 
a method which is  
a straightforward generalization of~\cite{PMH,BB}. Using a gauge
transformation the constant curvature 
connection can be brought into the form
\beqn
\nabla^i=\partial^i+i F^{ij}\sigma_j,   \label{CCcon}
\eeqn
where $F$ is an antisymmetric matrix. We  
define the constant curvature to be
\[
{\cal F}^{jk}_{(0)}=i\,[\nabla^j,\nabla^k]
\]
and using the commutation relations~(\ref{sigcom}) one can calculate
\[
{\cal F}_{(0)}=(2F+2\pi F\Theta F).
\]
In general, such a connection can only exist on a non-trivial bundle. 
One can introduce transition functions $\Omega_{i}$ such that the 
connection satisfies the twisted boundary conditions
\beqn
\nabla^i(\sigma_m+2\pi\delta^{j}_{m})=\Omega_j(\sigma_m) \nabla^i(\sigma_m) 
\Omega_j^{-1}(\sigma_m) \label{bcD}.
\eeqn
We can try to find solutions for the transition functions of the form 
\beqn
\Omega_j= e^{ i P^{jl}\sigma_l} W_j \label{omeg},
\eeqn
where $P$ is an arbitrary constant $d$-dimensional matrix 
and the $W_{i}$'s are constant, unitary $n$-dimensional matrices. 
The boundary conditions~(\ref{bcD}) imply the following equivalent relations
\[
P= (1+2\pi F \Theta)^{-1} 2\pi F = 2\pi F (1+\Theta 2\pi F)^{-1},
\]
\[
2\pi F = P(1-\Theta P)^{-1} = ( 1- P  \Theta)^{-1} P.
\]
Note that $P$ must be antisymmetric because of our gauge choice. 
Consistency of the transition functions 
of the bundle requires
\[
\Omega_j(\sigma_m +2\pi\delta^{i}_{m})\Omega_i(\sigma_m)=
\Omega_i(\sigma_m+2\pi\delta^{j}_{m})\Omega_j(\sigma_m),
\]
which is known in the mathematical literature 
as the cocycle condition. In our case the cocycle 
condition implies
\beqn
W_i W_j = e^{-2\pi i M^{ij}/n} W_j W_i, \label{WW}
\eeqn
where the antisymmetric matrix $M$ is given by
\[
M=n(2 P- P\Theta P).
\]
By taking the determinant of both sides of~(\ref{WW}) one finds 
that $M$ must have integer entries. In the classical case 
$M^{ij}$ corresponds to the value of the first Chern class
on the $(ij)$ two-cycle of the torus.
In the auxiliary Type IIA string 
theory, $M$ is interpreted as D2 brane winding. This interpretation 
remains true in the quantum case.

Let $q$ be the greatest common divisor of
$n$ and the nonvanishing entries of $M$
\[
q={\rm gcd}(n,M^{ij}).
\]
Next we define $\bn$ and $\bM$ which have relatively prime entries
\[
n=q\bn,~~M=q\bM.
\]
It is convenient to consider 
$W_i$'s which have block diagonal form with $q$  identical blocks 
along the diagonal
\[
 W_i =
\left(
\begin{array}{ccc}
\widetilde{W}_i &  &   \\
 & \ddots &    \\
 &  & \widetilde{W}_i
\end{array}
\right).
\] 
Here $\widetilde{W}_i$ are  $\bn$-dimensional matrices. 
Alternatively we can write this in tensor product notation
\[
W_i = I_q \otimes \widetilde{W}_i.
\]
The transition functions are also block diagonal and can be 
written
\beqn
\Omega_i = I_q \otimes \omega_i. \label{lomega}
\eeqn
To find explicit boundary conditions, following
\mbox{'t Hooft}~\cite{Hooft2}, we 
make the ansatz 
\beqn
\widetilde{W}_i=U^{a^i}V^{b^i},   \label{ansatz}
\eeqn
where $a^{i}$ and $b^{i}$ are integers and  $U$ and $V$ are 
the clock and shift matrices~\cite{Hooft,Hooft2} 
\[
U_{kl}= e^{2\pi i (k-1)/\bn} \delta_{k,l},~~V_{kl}= \delta_{k+1,l}
,~~k,l=1,\ldots, \bn,
\]
and the subscripts are identified with period $\bn$.
They satisfy
\[
UV=e^{-2\pi i /\bn}VU.
\]
Then~(\ref{WW}) leads to the following relation
\beqn
\bM^{ij}= (a^i b^j - b^i a^j) ~~{\rm mod}(\bn). \label{Mab}
\eeqn

For two or three dimensional tori, one can 
find integers $a^{i}$ and $b^{i}$ such that~(\ref{Mab}) holds for 
arbitrary $M$, as will be shown in Sections~\ref{T2} and~\ref{T3}. 
In higher dimensional cases the ansatz is 
not sufficiently general to describe arbitrary bundles.  
In particular, we can always 
perform a change of lattice basis such that the only nonvanishing
components of $M$ are $M^{d-1,d}=-M^{d,d-1}$, while in general,
an arbitrary antisymmetric matrix can not be brought into such a form. 
Furthermore,
for $d>3$, even in the commutative case, generic bundles do not admit 
connections with vanishing $su(n)$ constant curvature. A more general 
construction could be obtained by allowing for an arbitrary 
constant curvature connection.

\section{Adjoint Sections on Twisted Bundles} 
\label{Adj}

In this section we analyze the structure of adjoint sections on 
twisted bundles. The scalar and fermion fields are examples of such 
sections. We will also write the connection as a sum of a constant 
curvature connection $\nabla^i$, and a fluctuating part $A^i$
\[
D^i=\nabla^i-i A^i.
\]
Note that $A^i$\, is also an adjoint section. Since it is 
the difference between two connections it transforms covariantly under
gauge transformations. It should not be confused with a gauge potential.
Adjoint sections are $n$-dimensional matrices with entries which are elements
of the quantum plane algebra~(\ref{sigcom}) and obey the 
twisted boundary conditions
\beqn
\Psi(\sigma_i +2\pi\delta^j_i)=\Omega_j \Psi(\sigma_i)\Omega_j^{-1} .
\label{psibc}
\eeqn

Next we will try to find the general solution of~\rref{psibc} and 
write it in unconstrained form, reflecting the 
global properties of the bundle.
First consider the simpler example of a $U(n)$\, NCSYM on a trivial bundle
over a two-torus. Since $\Omega_i = 1$ we have
\[
\Psi =\sum_{a,b =1}^{n} E^{ab}\otimes [\sum_{i_1 i_2 \in {\bf Z}}
\Psi^{ab}_{i_1 i_2} U_1^{i_1} U_2^{i_2}],
\]
where $E^{ab}$ are $n$-dimensional matrices with one nonzero
entry, $(E^{ab})_{ij}=\delta_i^a\delta_j^b$, and 
$\Psi^{ab}_{{i_1}{i_2}}$ are $c$-numbers. In other words, each matrix
element of the adjoint section is an arbitrary function on the quantum torus. 
If we consider a twisted  $U(n)$ bundle
with  magnetic flux $m$, such that
$n$ and $m$ are relatively prime, one can show~\cite{CDS,PMH,BB} that
the adjoint sections have the expansion
\[
\Psi = \sum_{i_1 i_2 \in {\bf Z}}
\Psi_{i_1 i_2} Z_1^{i_1} Z_2^{i_2},
\] 
where now the coefficients $\Psi_{i_1 i_2}$ are $c$-numbers, and
$Z_i$ are $n$-dimensional matrices with noncommutative entries
satisfying
\[
Z_1 Z_2 = e^{2\pi i  \theta'} Z_2 Z_1.
\]
Thus the $Z_i$'s  satisfy the commutation relations of a
generators of the quantum torus. This shows that the set of sections
is isomorphic to the set of functions on a dual torus, and is very
similar to the 
set of adjoint sections of a $U(1)$ NCSYM theory.
For two and three dimensional adjoint bundles with
arbitrary magnetic fluxes, we will show that the general solution
takes the form
\beqn
\Psi = 
\sum_{ab=1}^q E^{ab} \otimes [
\sum_{i_1 i_2\ldots i_d \in {\bf Z}}
\Psi^{ab}_{i_1 i_2\ldots i_d} Z_1^{i_1} Z_2^{i_2} \ldots Z_d^{i_d} ],~~d=2,3.
\label{glory}
\eeqn 
Here $E^{ab}$ are $q$ dimensional. 

We begin by writing $\Psi$ in tensor notation
\[
\Psi(\sigma_i)= \sum_{a,b=1}^{q} E^{ab} \otimes \Psi^{ab}(\sigma_i),
\]
where  $\Psi^{ab}(\sigma_i)$
are $\bn$-dimensional matrices with noncommutative entries.
Imposing the boundary conditions~\rref{psibc} and 
using~\rref{lomega} we obtain
\beqn
\Psi^{ab}(\sigma_i +2\pi\delta^j_i)=\omega_j\, \Psi^{ab}(\sigma_i)\,
\omega_j^{-1}.
\label{psiomega}
\eeqn
A less restrictive but very convenient constraint is obtained by shifting
$\sigma_i$ by $2\pi \bn$  using~(\ref{psiomega})
\beqn
\Psi^{ab}(\sigma_{i}+2{\pi}\bn{\delta}^{j}_{i})=
\omega_{j}^{\bn}\Psi^{ab}(\sigma_{i})
\omega^{-\bn}_{j}.
\label{nshift}
\eeqn
In~(\ref{nshift}) all the matrix factors disappear since 
$U^{\bn}=V^{\bn}=1$. The $\sigma_i$ dependent exponential 
of~(\ref{omeg}) survives
and acts like a translation operator due to the commutation 
relations~(\ref{sigcom}). This implies the following periodicity relation  
\beqn
\Psi^{ab}(\sigma_i+2\pi \bn (Q^{-1})^{j}_{\,i}) 
=
\Psi^{ab}(\sigma_i), \label{per}
\eeqn
where
\[
Q^{-1}=1-P \Theta.
\]
Next we try to find  solutions of the form
\beqn
Z_i = e^{i\sigma_j Q^{j}_{\,k} N^{k}_{\,\,i}/ \bn}~ U^{s_i} V^{t_i},~~
i=1{\ldots}d.
\label{z}
\eeqn
Here  $s_j$ and $t_j$ are integers  and the exponent 
was chosen so that it is compatible with the 
constraint~(\ref{per}) if the matrix $N$ has integer entries. One can show 
that
$Z_i$ is compatible with the boundary conditions~(\ref{psiomega}) if
\beqn
N^{i}_{\,j}= (b^i s_j - a^i t_j)~~{\rm mod}(\bn),  \label{Nst}
\eeqn
where $a^{i}$ and $b^{i}$ are defined by~(\ref{Mab}). 
In the next two sections we will consider in detail the two and 
three dimensional cases, and  find $a^i$, $b^i$,
$s_j$ and $t_j$ such that~\rref{Mab} and~\rref{Nst} hold. 
Furthermore, for properly chosen integers $a^i$, $b^i$,
$s_j$ and $t_j$,
one can show that an arbitrary adjoint section can be
expanded in terms of the $Z_i$'s as in~(\ref{glory}). For a proof of
this statement in two dimensions see~\cite{BB}.
It is convenient to define another matrix which
will be used shortly,
\beqn
L_{ij}= (s_i t_j - t_i s_j)~~{\rm mod}(\bn). \label{l}
\eeqn

In the  remainder of this section we will calculate the 
commutation relations satisfied by the  $Z_i$'s and the constant 
curvature connection~\rref{CCcon}. Using their explicit form~(\ref{z}) 
we find,
after some matrix algebra, 
\beqn
Z_i Z_j = e^{2\pi i \Theta'_{ij}}~Z_j Z_i  \label{ZZ}
\eeqn
where
\beqn
\Theta'= \bn^{-2} N^{T} Q^{T} \Theta Q N -\bn^{-1} L. \label{uglythet}
\eeqn
From~\rref{ZZ} we see that the algebra generated by the $Z_i$'s
is the algebra of functions on the quantum torus
with deformation parameters given by $\Theta'$. 
After some further matrix algebra and  using the following identities, 
\begin{eqnarray}
Q&=& 1+ 2\pi F \Theta,  \nonumber \\
Q^{2} & =& 1 + 2\pi  {\cal F}_{(0)}\Theta = (1- \bM\Theta/\bn)^{-1},
 \nonumber \\
 Q^T\Theta&=&\Theta Q~~~, \nonumber 
\end{eqnarray}
we can rewrite $\Theta'$ as a fractional transformation
\beqn
\Theta' = \Lambda(\Theta) \df ({\cal A}\Theta +{\cal B})
({\cal C}\Theta+{\cal D})^{-1} \label{lamthet}.
\eeqn
Here
\[
\Lambda=
\left(
\begin{array}{cc}
{\cal A} & {\cal B} \\
{\cal C} &{\cal  D}
\end{array}
\right),
\]
and the $d$-dimensional block matrices are given by
\beqn
{\cal A}=\bn^{-1}(N^{T}+LN^{-1}\bM),~ 
{\cal B}= -LN^{-1} ,~
{\cal C}=-N^{-1}\bM ,~
{\cal D}= \bn N^{-1}. \label{abcd}
\eeqn
One can check that
\beqn
{\cal A}^{T}{\cal D}+{\cal C}^{T}{\cal B}=1,~{\cal A}^{T}{\cal C}
+{\cal C}^{T}{\cal A}=0
,~{\cal B}^{T}{\cal D}+{\cal D}^{T}{\cal B}=0,
\label{Od}
\eeqn
and thus $\Lambda$ is an element of $O(d,d\,|{\bf R})$, i.e. it
satisfies 
\[
\Lambda^{T} J \Lambda = J,
\]
where
\beqn
J=
\left(
\begin{array}{cc}
0 & I_d \\
I_d & 0
\end{array}
\right).  \label{Jmetric}
\eeqn
In the two and three dimensional examples that we will study later,
$\Lambda$ is in fact an element of $SO(d,d\,|{\bf Z})$.
This is the subgroup with  determinant one and integer valued entries 
in the basis where the metric is given by~(\ref{Jmetric}). 
The Weyl spinor representations of $SO(d,d\,|{\bf Z})$ are also integral,
that is the representation matrices have integer entries. We
prove  this statement, which is implicit in papers discussing
T-duality of Type II string theory,
in the appendix.
Since the spinor representation of $SO(d,d\,|{\bf Z})$ will be used
extensively  in the following sections we recall that 
the vector and 
spinor representations are related  by
\beqn
{\cal S}^{-1} \gamma_{s}~{\cal S}= \Lambda_{s}^{~p} ~\gamma_{p},
\label{Sgam}
\eeqn
and the gamma matrices satisfy
\beqn
\{\gamma_s,\gamma_p\}=2 J_{sp}.  \label{gammacom}
\eeqn

Finally, one can show by direct calculation 
that the commutation relations of the
constant curvature connection and  $Z_i$  
have the form
\beqn
[\nabla^i,Z_j]= i H^{i}_{\,j} Z_{j}, \label{dzhz}
\eeqn
where there is no sum over $j$ and
$ H = (\bn-\bM\Theta)^{-1} N$. 
Note that $H$ can also be  written in terms of $\Theta$ and 
some of the block components 
of $\Lambda$
\beqn
H^{-1} = {\cal C} \Theta + {\cal D}. \label{nice}
\eeqn
Finally, we present some identities,  which will be useful in later
sections
\beqn
H  = \bn^{-1} Q^{2} N, ~~
\det(H)  =  \bn^{-2} \det(Q^{2}), \label{uid}  
\eeqn
\[
\det(Q^{2})  =  (1- \frac{{\rm tr} (\bM\Theta)}{2\bn})^{-2}. 
\]
Note that all the previous relations are valid for tori of 
arbitrary dimension provided
we work on the bundles discussed in Section~\ref{QB}.

\section{Two Dimensional Solution}
\label{T2}

Although the twisted two dimensional case has been discussed extensively
in the literature~\cite{CDS,AAS,PMH,BB}, we review 
it here in a form that readily admits generalization to higher
dimensional compactifications. 

In the two dimensional case 
the antisymmetric matrices $\Theta$ and $M$ have the form
\[
\Theta=\left(
\begin{array}{cc}
0 & \theta  \\
-\theta & 0
\end{array}
\right),~
M=\left(
\begin{array}{cc}
0 & m  \\
-m & 0
\end{array}
\right)
\]
where $\theta$\, is the deformation parameter and $m$\, is the 
magnetic flux, which is interpreted
as the number of D2 branes wrapping the two-torus.

One can verify that the integers
\[
(a^i) =(\bm,0),~(b^i)= (0,1),
\]
where $n=q\bn$ and $m=q\bm$, satisfy~(\ref{Mab}). 
Then choosing $s_i= (0,1)$ and $t_i =(b,0)$, where
$b$ is an integer such that $a\bn-b\bm=1$, we have $N=I_2$.
One can now use~(\ref{l}) and~(\ref{abcd}) to find
\beqn
\Lambda =\left(
\matcc
a  I_2 & b \varepsilon \\
-\bm \varepsilon & \bn I_2
\emat
\right), \label{lamtwo}
\eeqn
where $\varepsilon$ is a two dimensional matrix with the only
nonvanishing  
entries given by 
$\varepsilon_{12}= -\varepsilon_{21}=1$.
Group elements of the form above are in an $SL(2,{\bf Z})$ subgroup of
$SO(2,2\,|{\bf Z})$. This subgroup is isomorphic with one of
the Weyl spinor representations of $SO(2,2\,|{\bf Z})$. This feature
is not generic for higher dimensional compactifications and reflects
the fact that $SO(2,2\,|{\bf Z}) \sim SL(2,{\bf Z})\times 
SL(2,{\bf Z})$, so that it is not simple.

The algebra of the $Z_i$'s is then determined by
$\Theta'$ which is given by the fractional transformation~\rref{lamthet}.
In two dimensions, the $SO(2,2\,|{\bf Z})$ fractional 
transformation~(\ref{lamthet}) can 
also be written in the more familiar form, used in~\cite{CDS,PMH},
as a $SL(2,{\bf Z})$ fractional 
transformation acting on  $\theta$
\beqn
\theta'=\frac{a\theta+b}{\bm\theta+ \bn}. \label{thetaprime}
\eeqn
One can also check that the other $SL(2,{\bf Z})$ subgroup, 
made of elements of the form
\[
\left(
\matcc
R & 0 \\
0   & (R^{T})^{-1}
\emat
\right),
\]
acts trivially on $\theta$. This subgroup is generalized to
$SL(d,{\bf Z})$ in compactifications on a $d$-dimensional torus, 
and will play in important role later, but only 
for the two dimensional compactification 
it leaves $\Theta$ invariant.
The $Z_i$'s then obey the following algebra

\[
Z_1 Z_2 = e^{2\pi i \theta'}Z_2 Z_1.
\]

As we will see shortly,
the rank of the gauge group and the magnetic flux transform 
in an integral Weyl spinor representation of $SO(2,2\,|{\bf Z})$. 
Using the creation and annihilation operators introduced in the
appendix we can write such a spinor as
\beqn
n|0\rangle+ma^{\dagger}_{1}a^{\dagger}_{2}|0\rangle. \label{Focktwo}
\eeqn
Using~\rref{Sgam} one can show that the spinor representation
of~\rref{lamtwo}  transforms the above state into $q|0\rangle$. 
In the Weyl basis we can write the action as 
\[
\left( \begin{array}{c} q \\ 0 \end{array} \right)=
S \left( \begin{array}{c} n \\ m \end{array} \right),
\]
where
\[
S= \left(
\matcc 
a  &  -b \\
-\bm  &  \bn
\emat
\right).
\]
In Section~\ref{DSYM} we will show, employing the expansion of the adjoint
section in terms of the $Z_i$ generators~(\ref{glory}), how to rewrite 
the original
$U(n)$\, NCSYM action on a twisted bundle as a $U(q)$\, NCSYM action
on a trivial quantum bundle over a 
torus with deformation parameter $\Theta'$. The $SL(2,{\bf Z})$ 
transformation, which relates the deformation parameters 
and the spinors~(\ref{Focktwo}) of these two  NCSYM,
can then be interpreted as a duality transformation 
inherited from T-duality  of 
Type II string theory. This can be seen as follows. The rank and the bundle 
of the NCSYM theory determine the D brane charges in string theory.
These charges transform in a chiral spinor representation of the target space 
duality group~\cite{EdW}. Given $n$ and $m$ with greatest common divisor
$q$,
one can perform a T-duality transformation which takes the original
D brane configuration into $q$ D0 branes. 

Of course the metric and 
antisymmetric tensor also transform under this duality, and in
the proper limit, which we will explain in detail later, the
antisymmetric tensor $B$ transforms separately by fractional transformation 
just as in~(\ref{lamthet}).  Since the parameters $\Theta_{ij}$ of the 
NCSYM theory are identified 
with $B_{ij}$, the background expectation
value of the NS antisymmetric tensor 
of the compactified auxiliary string theory, the
expected transformation under target space dualities is
(\ref{lamthet}).

\section{Three Dimensional Solution}
\label{T3}

The three dimensional case will be solved by first 
performing an $SL(3,{\bf Z})$ transformation $R$
to bring $M$ in canonical 
form\footnote{It is always possible to bring an antisymmetric matrix in
canonical form using $SL(3,{\bf R})$ but here we need to do this 
using an integral matrix.}
\beqn
M=R M^0 R^{T},\label{canonic}
\eeqn
where
\beqn
M^0 =
\left(
\begin{array}{ccc}
0 & 0 & 0 \\
0 & 0 & m \\
0 &-m & 0
\end{array}
\right).  \label{M0}
\eeqn
While it is always possible to find such a transformation,~\rref{canonic}
does not define it uniquely.
We will first find the solution corresponding to $M^{0}$, and then
obtain the general solution by using such an $R$.

First note that $M^{0}$ corresponds to a background magnetic 
field  with flux only through the (23) plane, which suggests that 
the solution 
should closely resemble the two dimensional one.
As before,
\[
(a^i_0)=(0,\bm,0),~(b^i_0)=(0,0,1)
\]
satisfy~(\ref{Mab}).
Similarly if we set  
\beqn
(s_i^0)=(0,0,1),~(t_i^0)=(0,b,0), \label{sandt}
\eeqn
we can satisfy~\rref{Nst} with the $N^{0}$ matrix given by
\[
N^0 =
\left(
\begin{array}{ccc}
\bn & 0 & 0 \\
0 & 1 & 0 \\
0 & 0 & 1
\end{array}
\right).
\]
The diagonal entries of $N^{0}$ divided by $\bn$ have the interpretation of
wave numbers. Thus we see that twisting the boundary conditions allows for
fractional wave numbers in the second and third directions.
Using~(\ref{sandt}) we find 
\[
L^0 =
\left(
\begin{array}{ccc}
0 & 0 & 0 \\
0 & 0 &-b \\
0 & b & 0
\end{array}
\right).
\]
One can now use~(\ref{abcd}) to find the $SO(3,3\,|{\bf Z})$ 
matrix
\beqn
\Lambda^{0}=
\left(
\begin{array}{cccccc}
1 & 0 & 0 & 0 & 0 & 0 \\
0 & a & 0 & 0 & 0 & b \\
0 & 0 & a & 0 &-b & 0 \\
0 & 0 & 0 & 1 & 0 & 0 \\
0 & 0 &-\bm & 0 & \bn & 0 \\
0 & \bm & 0 & 0 & 0 & \bn \\
\end{array}
\right).
\label{Lam0}
\eeqn
Everything so far is just as in the two dimensional case.
Note however that in general  $\Theta$ will not be in canonical form,
that is, 
it will not have a form similar to~\rref{M0}.

We can now write the general solution for an arbitrary $M$ as
\[
a^i = R^{i}_{\,j} ~a^j_0,~~
b^i = R^{i}_{\,j} ~b^j_0,
\]
\[
s_i = s_j^0,~~
t_i = t_j^0,
\]
\[
N=RN^{0},
\]
\beqn
\Lambda
=
\Lambda^{0}
\left(
\begin{array}{cc}
R^{T} & 0 \\
0 & R^{-1}
\end{array}
\right).
\label{Lam}
\eeqn
Just as in the two dimensional case we can find, using~\rref{Sgam},
the Weyl spinor representation
matrices corresponding to~\rref{Lam0} and~\rref{Lam}
\[
S^{0}=
\left(
\begin{array}{cccc}
a & -b & 0 & 0 \\
-\bm & \bn & 0 & 0 \\
0 & 0 & 1 & 0 \\
0 & 0 & 0 & 1
\end{array}
\right)
,~~
\]
\[
S= S^{0} 
\left(
\begin{array}{cc}
1 & 0 \\
0 & R^{T}
\end{array}
\right).
\]
The rank of the group and the magnetic flux matrix $M$ define a  
state in the  Weyl spinor Fock space
\[
n|0\rangle+\frac{1}{2}M^{ij}a^{\dagger}_{i}a^{\dagger}_{j}|0\rangle.
\]
Now one can check that $S$ acts on this spinor as
\[
\left(
\begin{array}{c}
q\\
0\\
0\\
0\\
\end{array}
\right)=S
\left(
\begin{array}{c}
n\\
M^{23}\\
M^{31}\\
M^{12}\\
\end{array}
\right).
\]
As we will see later this can be used to relate the original theory 
to a $U(q)$ theory on a trivial bundle.
In the appendix we show that the Weyl spinor representation of
$SO(3,3\,|{\bf Z})$ is in fact isomorphic to $SL(4,{\bf Z})$.
In this case, in the auxiliary Type IIA string theory,
the D0 and D2 branes form $q$ 
bound states, and the transformation above corresponds to
a T-duality transformation
that maps the original D brane configuration into a $q$ D0 branes.

\section{Noncommutative Super Yang-Mills Action}
\label{NCSYM}

We are now almost ready to write the noncommutative Super Yang-Mills action,
but first we need to understand how to perform integration on a noncommutative
torus. In the classical case the integral is a linear map that associates 
to a function its zero mode Fourier coefficient.
Similarly for an element of ${\cal A}_{\Theta}$ of the form 
$a=\sum  a_{i_1 i_2 \ldots i_d} U_1^{i_1} U_2^{i_2}\ldots
U_d^{i_d}$\, we define the integral as
\beqn
\int  d^{d} \sigma\,  a \df {(2\pi)^d}\, a_{00\ldots 0}.\label{integra}
\eeqn
One can check that this definition has all the desirable properties
of the classical integral, such as linearity and translation 
invariance in $\sigma_i$. For definiteness, 
in the remainder of this section we will discuss
the three dimensional case.

When twisted  $U(n)$ theories are considered, it was found 
in~\cite{CDS,PMH} that the integral must be normalized in a
particular way to find a duality invariant spectrum. 
The normalization can also be obtained directly
as the Jacobian of a change of integration variables.
Note that the integrand, 
which is the trace of an adjoint section, obeys the following periodicity
\[
{\rm tr} \Psi(\sigma_i) ={\rm tr} \Psi(\sigma_i+ 2\pi (Q^{-1})_{\,i}^{j}).
\]
Since ${\rm tr} \Psi(\sigma_i)$ does not have periodicity $2\pi$
in $\sigma_i$ it can not be expanded in terms of the $U_i$\,
variables. One can define new variables 
$\widehat{\sigma}_i=\sigma_j Q^{j}_{\,k} R^{k}_{\,i}$ and $\widehat{U}_i=
e^{\widehat{\sigma}_i}$, where $R$ is an arbitrary $SL(3,{\bf Z})$ 
transformation. In the following sections we will take $R$ to be the
matrix that brings $M$ into canonical form~\rref{canonic}.
Then
\beqn
\int d^{3}\sigma\, {\rm tr}\, \Psi(\sigma) =
\int  d^{3}\widehat{\sigma}\,| \det( Q^{-1})|\, 
{\rm tr}\, \Psi(\widehat{\sigma}Q^{-1}),  \label{hatsig}
\eeqn
where $\det( Q^{-1})$ is the Jacobian of the coordinate transformation,
and the second integral can now be performed as discussed above, 
since the integrand has an expansion
in terms of the $\widehat{U}_i$\ variables.
Using the expansion~(\ref{glory}) of $\Psi$ we obtain
\[
\int d^{3}\sigma\, {\rm tr}\, \Psi(\sigma) =
(2\pi)^3 \, \bn\,|\det(Q^{-1})|\,\sum_{a=1}^{q} \Psi^{aa}_{000}.
\]

The Super Yang-Mills action on a noncommutative three-torus is given by
\[
{\cal S}^{U(n)} = \frac{1}{g^{2}_{SYM}} \int dt
\int d^{3} \sigma~\sqrt{\det(G^{ij})}
~{\rm tr} \left(
\frac{1}{2} G_{ij} {\cal F}^{0i}  {\cal F}^{0j} -  \right.
\]
\beqn
\frac{1}{4} G_{ij} G_{kl} ({\cal F}^{ik}-{\cal F}^{ik}_{(0)})
({\cal F}^{jl}-{\cal F}^{jl}_{(0)}) +
\label{Unaction}
\eeqn
\[
\frac{1}{2} \sum_{a} \dot{X}^a  \dot{X}^a -
\frac{1}{2} \sum_{a} G_{ij} [D^i,X^a] [D^j,X^a]+
\]
\[
\left.
\frac{1}{4} \sum_{a,b} [X^a,X^b] [X^a,X^b]
+{\rm fermions}\right),
\]
where ${\cal F}^{ij} = i\,[D^{i},D^{j}]$ 
and ${\cal F}_{0}^{ij} = i\,[\nabla^{i},\nabla^{j}]$.
We have subtracted the constant part of the field strength in the second
line of equation~\rref{Unaction}. This is equivalent to adding a 
constant to the Lagrangian, or equivalently to the Hamiltonian, and  has the
effect of setting the vacuum energy to zero. 

For the  compactification of the auxiliary Type IIA string theory
without wrapped D2 branes,
the above action can be obtained directly from the Matrix
action. 
One has to show that the trace over infinite dimensional 
matrices reduces to a finite dimensional trace and an integral.
A formal argument for the commutative case was given in~\cite{WTc} 
and discussed in detail in~\cite{WT}. The same argument extends to the 
noncommutative case. 
A brief argument was given in~\cite{CDS} 
showing how to extend this construction
when there are D2 branes wrapped on the torus 
in the auxiliary Type IIA string theory,
corresponding to magnetic fluxes in the NCSYM gauge theory. 
Here we will just make the assumption that the 
NCSYM action is independent of the D2 brane charges and that adding
D2 branes only results in changing the quantum adjoint bundle.
We will provide evidence for
this in the final section of the paper.

\section{$SO(3,3\,|{\bf Z})$ Duality of Super Yang-Mills}
\label{DSYM}

In this section we start with the  $U(n)$ NCSYM action~\rref{Unaction} 
on a twisted quantum bundle with magnetic fluxes $M$ and deformation 
parameter $\Theta$, and we show that
after a sequence of field redefinitions it can be rewritten as a  
$U(q)$ NCSYM action on a trivial bundle  over a
quantum torus with deformation parameter $\Theta'$.

Using the matrix $H$ defined in~\rref{dzhz} we make the following
constant curvature and field
redefinitions
\[
\widehat{\nabla}^i \df (H^{-1})^{i}_{\,j} \nabla^j,
~\widehat{A}^i \df (H^{-1})^{i}_{\,j} A^j,
\]
\[
~\widehat{D}^i \df (H^{-1})^{i}_{\,j} D^j,
\]
\[
{\cal \widehat{F}}^{kl}=
[\widehat{\nabla}^k,\widehat{A}^l]-[\widehat{\nabla}^l,\widehat{A}^k]-
i[\widehat{A}^k,\widehat{A}^l].
\]
In terms of the new variables the commutator of the constant curvature 
connection and the $Z_i$'s
takes the simple form,
\[
[\widehat{\nabla}^i, Z_j] = i\,\delta^i_j Z_j,
\]
and the curvature can be expressed as
\[
{\cal F}^{ij} = {\cal F}_{(0)}^{ij} +
H^{i}_{\,k} H^{j}_{\,\,l}\,
{\cal \widehat{F}}^{kl}.
\]
One can now rewrite the action in terms of the hatted variables 
and perform the change of coordinates~\rref{hatsig}
\[
{\cal S}^{U(n)} = \frac{1}{{g'}^{2}_{SYM}} \int dt
\int d^{3} \widehat{\sigma}~\sqrt{\det(G'^{ij})}
\frac{1}{\bn}~{\rm tr} \left(
\frac{1}{2} G'_{ij} {\cal \widehat{F}}^{0i}  {\cal \widehat{F}}^{0j} - 
\right.
\]
\[
\frac{1}{4} G'_{ij} G'_{kl} {\cal \widehat{F}}^{ik}
{\cal \widehat{F}}^{jl} +
\]
\[
\frac{1}{2} \sum_{a} \dot{X}^a  \dot{X}^a-
\frac{1}{2} \sum_{a} G'_{ij} [\widehat{D}^i,X^a] [\widehat{D}^j,X^a]+
\]
\[
\left.
\frac{1}{4} \sum_{a,b} [X^a,X^b] [X^a,X^b]
+{\rm fermions}\right).
\]
We have introduced a new gauge coupling and metric given by
\beqn
{g'}_{SYM}^{2}= \bn\,|\det( Q^{-1} )|~g_{SYM}^{2} \label{Gcoupling}
\eeqn
\beqn
G'^{ij}= (H^{-1})^{i}_{\,k} (H^{-1})^{j}_{\,\,l} \, G^{kl}
\label{metric}
\eeqn
and used~\rref{uid} to make these substitutions.

Next we introduce primed variables  $\sigma'_i$, 
$U'_{i}$ and  partial derivatives $\partial'^{i}$ satisfying
\[
[\sigma'_i,\sigma'_j] = -2\pi i \Theta'_{ij},
\]
\[
[\partial'^i,\sigma'_j] = \delta^i_j,~~
[\partial'^{i},\partial'^{j}]=0,
\]
\[
U'_{i} \df e^{i\sigma'_i},
\]
\[
U'_{i} U'_{j}= e^{2\pi i \Theta'_{ij}} U'_{j} U'_{j}.
\]
Comparing the algebra satisfied by $Z_i$ and $\widehat{\nabla}^i$ on one hand
and $U'_i$ and $\partial'_i$ on the other, we see that all the commutation
relations are the same except that the $\widehat{\nabla}^i$'s do not commute
while the $\partial'^i$'s do. The dynamical variables of the theory are
the $c$\,-number coefficients appearing in the expansion~\rref{glory} 
of the adjoint sections in terms of $Z_i$'s. 
Since in the action, the constant curvature covariant derivatives 
only appear in commutators with the $Z_i$'s and not with each other,
substituting  $U'_i$ and $\partial'^i$ for $Z_i$ and $\widehat{\nabla}^i$ 
leaves the dynamics invariant. A similar construction was also considered
in~\cite{AS2}.
The integral and trace of the $U(n)$ theory can be translated to a
$U(q)$ integral using the definition of the integral~\rref{integra}
\[
\int d^{3}\widehat{\sigma}\, \frac{1}{\bn} {\rm tr}\,
 \Psi(Z_i)=
\int d^{3}\sigma'  \, {\rm tr}_q \, \Psi(U'_i)=
(2\pi)^3 \, \sum_{a=1}^{q}\Psi^{aa}_{000}.
\]
Making these substitutions we obtain the $U(q)$ action
\[
{\cal S}^{U(q)} = \frac{1}{g'^{2}_{SYM}} \int dt
\int d^{3} \sigma'~\sqrt{\det(G'^{ij})}
~ {\rm tr}_q\left(
\frac{1}{2} G'_{ij} {\cal F'}^{0i}  {\cal F'}^{0j} -  \right.
\]
\[
\frac{1}{4} G'_{ij} G'_{kl} {\cal F'}^{ik}
{\cal F'}^{jl}+
\]
\[
\frac{1}{2} \sum_{a} \dot{X}^a  \dot{X}^a-
\frac{1}{2} \sum_{a} G'_{ij} [D'^i,X^a] [D'^j,X^a]+
\]
\[
\left.
\frac{1}{4} \sum_{a,b} [X^a,X^b] [X^a,X^b]
+{\rm fermions}\right),
\]
where
\[
D'^{i} \df  \partial'^{i} -i A'^{i},~
{\cal F'}^{ij} =i\, [D'^{i},D'^{j}]
\]
are the  $U(q)$ connection and  curvature. Thus we have shown that the
original $U(n)$ theory is equivalent to a $U(q)$ NCSYM theory with
gauge coupling given by~\rref{Gcoupling} and 
defined on a trivial adjoint bundle over a 
noncommutative torus with deformation parameter
$\Theta'$ and metric given by~\rref{metric}.

In general two NCSYM theories are dual to each other if there exists an
element $\Lambda$ of $SO(3,3\,|{\bf Z})$  with 
Weyl spinor representation matrix $S$, such that their defining parameters
are related as follows
\beqn
\bar{\Theta}= ({\cal A}\Theta +{\cal B})({\cal C}\Theta+{\cal D})^{-1},
\label{Theta2}
\eeqn
\beqn
\left(
\begin{array}{c}
\bar{n}\\
\bar{M}^{23}\\
\bar{M}^{31}\\
\bar{M}^{12}\\
\end{array}
\right)=S
\left(
\begin{array}{c}
n\\
M^{23}\\
M^{31}\\
M^{12}\\
\end{array}
\right),
\label{n2}
\eeqn
\beqn
 \bar{G}^{ij} = 
({\cal C}\Theta +{\cal D})^{i}_{\,k} 
({\cal C}\Theta +{\cal D})^{j}_{\,\,l} \, G^{kl},
\label{G2}
\eeqn
\beqn
\bar{g}_{SYM}^{2}= 
\sqrt{\,|\det({\cal C}\Theta+{\cal D})|}~g_{SYM}^{2},
\label{gsym2}
\eeqn
where we used~\rref{nice} in the last two equations.
While $\Theta$ in~\rref{Theta2} and the rank and magnetic flux 
numbers in~\rref{n2} transform separately and the duality group action 
can be seen explicitly, the transformation of the 
gauge coupling and the metric also depends on
$\Theta$. 
Note that ${\cal C}\Theta+{\cal D}$ satisfies a group property.
If  $\Lambda_3=\Lambda_2 \Lambda_1$ and $\Theta'=\Lambda_1(\Theta)$ then
\beqn
{\cal C}_3\Theta+{\cal D}_3=
({\cal C}_2\Theta'+{\cal D}_2 )({\cal C}_1\Theta+{\cal D}_1).
\label{groupP}
\eeqn
For a nonvanishing $\bar{n}$ we remove
the sign ambiguity that exists
when we try to associate to a $\SOS$ transformation its spinor
representation matrix, by requiring that $\bar{n}$ is positive.
Strictly speaking, one should not consider duality transformations 
for which $\bar{n}$ vanishes since in this case the description 
in terms of gauge theories 
becomes singular.

\section{Target Space Duality}
\label{TDuality}

Next we show that the $\SOS$ duality discussed in the previous section
is the realization in  NCSYM gauge theories 
of T-duality in  the auxiliary Type IIA string theory.
This relation is described by the following diagram.
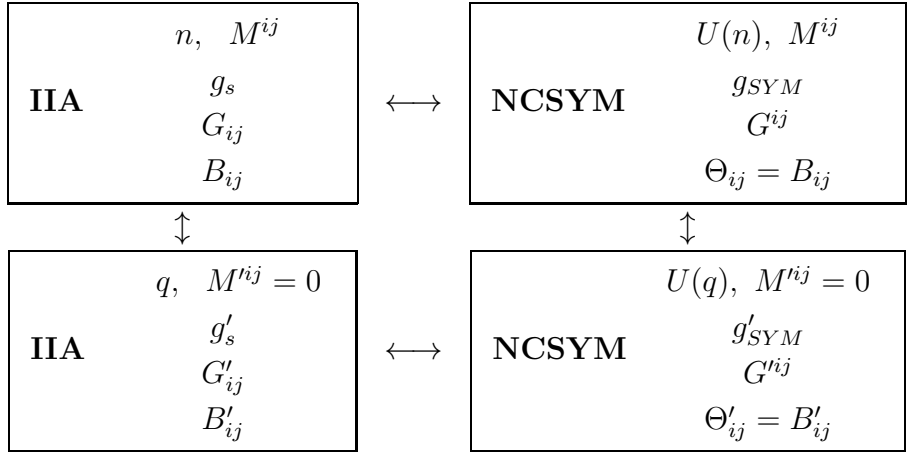
\thicklines
\[\]\[
\begin{array}{ccc}
  	 \fbox{$\begin{array}{cc}{\bf IIA} &
	\begin{array}{c}
		~~~~~n,~~ M^{ij}~~~~\\ 
		g_s\\ 
		G_{ij}\\
		B_{ij}
	\end{array}\\
	\end{array}$}
& \longleftrightarrow & 
	\fbox{$\begin{array}{cc}{\bf NCSYM} &
	\begin{array}{c} 
		~~~U(n),~ M^{ij}~~~\\
		g_{SYM}\\
		 G^{ij} \\
		\Theta_{ij}=B_{ij}
		\end{array}
	\end{array}$} \\
\updownarrow &    & \updownarrow    \\
	 \fbox{$\begin{array}{cc}{\bf IIA} &
	\begin{array}{c}
		~~~q,~~ M'^{ij}=0\\ 
		g'_s\\ 
		G'_{ij}\\
		B'_{ij}
	\end{array}\\
	\end{array}$}
& \longleftrightarrow & 
	\fbox{$\begin{array}{cc}{\bf NCSYM} &
	\begin{array}{c} 
		U(q), ~M'^{ij}=0\\
		g'_{SYM}\\
		 G'^{ij} \\
		\Theta'_{ij}=B'_{ij}
		\end{array}
	\end{array}$} \\
\end{array}
\]
\[\]
\[\]

The right side of the diagram  
shows the equivalence described in Section~\rref{DSYM}. The horizontal
arrows represent the Connes, Douglas and Schwarz conjecture~\cite{CDS}.
The left side of the diagram contains the string coupling, 
\mbox{D brane} charges, and compactification moduli
of the two auxiliary Type IIA string theories corresponding to the
NCSYM's on the right.
The additional moduli corresponding to  Ramond-Ramond backgrounds were 
set to zero in this paper and will be considered separately in~\cite{BM}. 
Note that the NCSYM metric is the inverse of the Type IIA metric
as indicated by the index position, the
deformation parameter equals the NS antisymmetric tensor, and the rank
and magnetic flux numbers translate into D0 brane number and D2 brane
winding. Finally the SYM and string coupling are related by~\rref{ggg}.

In the remainder of this section we will calculate the relation between the 
parameters of the two auxiliary Type IIA string theories. First we 
describe how the metric, antisymmetric tensor and the string coupling
transform under an arbitrary T-duality transformation, 
and then we take the limit 
\beqn
\alpha' \rightarrow 0,~~G_{ij} \rightarrow 0,
\label{limit}
\eeqn
keeping  $\alpha'^{-2} G_{ij}$ constant.
This is the limit proposed by Seiberg and Sen~\cite{NS,AS} and 
briefly discussed in the introduction.
However, in this limit the auxiliary 
Type IIA string metric vanishes. Instead we will 
calculate directly 
the inverse metric of the NCSYM theory which, after including  
factors of $\alpha'$, is given by 
$\alpha'^{-2}G_{ij}$. 

Under the T-duality group $SO(d,d\,|{\bf Z})$ the metric and NS antisymmetric 
tensor\footnote{We hope there is no confusion between $B$, denoting
the NS tensor, and ${\cal B}$ which is the upper right block of $\Lambda$.} 
transform together by fractional transformations~\cite{GPR}
\beqn
G'+B' = ({\cal A}(G+B) +{\cal B})({\cal C}(G+B)+{\cal D})^{-1}.
\label{GplusB}
\eeqn
Using the identification between $\Theta$ and $B$ we have 
$H^{-1}={\cal C}B+{\cal D}$. Then, after some matrix algebra, we can 
write the symmetric and antisymmetric part of~\rref{GplusB}
as
\begin{eqnarray}
G'&=& H^{T} G (1-(H{\cal C}G)^{2})^{-1} H \label{Gp},  \\
B'&=&({\cal A}B +{\cal B})({\cal C}B+{\cal D})^{-1}-
H^{T} G H{\cal C} G(1-(H{\cal C}G)^{2})^{-1}H.  \label{Bp}
\end{eqnarray}
To derive this we used the fact that $H{\cal C}$ is antisymmetric.
This can be shown using
\[
({\cal C}B+{\cal D})^{-1}=
({\cal A}-({\cal A}B+{\cal B})({\cal C}B+{\cal D})^{-1}{\cal C})^{T},
\]
which follows from~\rref{Od}. 
Note that~\rref{Gp} and~\rref{Bp} have simple expansions in $G$.
For an elementary T-duality in the $x^{1}$ direction 
the string coupling constant transforms as
\beqn
g_{S}'=g_{S}  G_{11}^{-1/2}.
\label{gT}
\eeqn

Taking the limit~\rref{limit} in~\rref{Bp} we can see that the 
antisymmetric tensor itself transforms by fractional 
transformation\footnote{This is consistent with the fact that
the action by fractional transformations preserves the antisymmetry
of the  matrices.}
\beqn
B' = ({\cal A}B +{\cal B})({\cal C}B+{\cal D})^{-1}.
\label{Bprime}
\eeqn

To find the duality transformation of the metric, we reinstate factors of
$\alpha'$ in~\rref{Gp} since the $SO(3,3\,|{\bf Z})$ transformations 
are defined to act on dimensionless fields. 
%Then 
%\[
%\alpha'^{-2}G'=\alpha'^{-1} \left[ \Lambda(\alpha'^{-1}G+ B)\right]_S,
%\]
%where the subscript $S$ indicates that only the symmetric part should 
%be taken.
Now we can take the limit~\rref{limit} and
 to first order in the dimensionless 
metric $\alpha'^{-1} G_{ij}$ we have
\beqn
\alpha'^{-2}G'= 
({\cal C} B + {\cal D})^{-T}(\alpha'^{-2}G)({\cal C}B + {\cal D})^{-1} .
\label{Gprime}
\eeqn
If we make the identification $B=\Theta$, we recognize above
the $H$
matrix defined in~(\ref{nice}). 

Finally using~(\ref{gT}), we can also calculate how the string
coupling transforms under duality. 
It  was shown in~\cite{AS1} that 
the $SO(d,d\,|{\bf Z})$ group is generated by a set of simple elements.
These are written explicitly in the appendix. For each of these generators one 
can check using~(\ref{gT}) that the string coupling transforms as
\beqn
g_{S}'= g_{S}\,   |\det({\cal C}B + {\cal D})|^{-1/2} .
\label{gstring}
\eeqn
In fact~\rref{gstring} is true for an arbitrary transformation
because ${\cal C}B + {\cal D}$ satisfies the group property~\rref{groupP}.

Comparing the T-duality relations~\rref{Bprime},~\rref{Gprime} 
and~\rref{gstring} with the NCSYM duality 
relations~\rref{Theta2},~\rref{G2} 
and~\rref{gsym2}, using~\rref{ggg} to relate the string and 
gauge couplings, we see that indeed the two dualities coincide.

\section*{Acknowledgments}
This work was supported in part by 
the Director, Office of Energy Research, Office of High Energy and Nuclear
Physics, Division of High Energy Physics of the U.S. Department of Energy
under Contract DE-AC03-76SF00098 and in part by the National Science 
Foundation under grant PHY-95-14797.

\appendix
\section*{Appendix}
In the first part of this appendix we show that the Weyl spinor
representations of $SO(d,d\,|{\bf Z})$ are integral, i.e. have matrix
elements which are integers. In the final part, we show that for
$d=3$ the Weyl spinor representation is in fact isomorphic to
$SL(4,{\bf Z})$. 

The gamma matrices obeying~(\ref{gammacom}),  where the metric has the 
form~(\ref{Jmetric}), are already, up to normalization, the
standard creation and annihilation operators used to generate
the Fock space for Dirac spinors in the Weyl basis. These are defined as
\[
a_i^{\dagger}=\gamma_i/\sqrt{2},~~a_i=\gamma_{d+i}/\sqrt{2}
\]
and satisfy the canonical anti-commutation relations
\[
\{a_{i},a_{j}^{\dagger}\} = \delta_{ij},~~
 \{a_{i},a_{j}\} = \{a_{i}^{\dagger},a_{j}^{\dagger}\} = 0,~~
i,j=1,\ldots,d.
\]
As usual, the Dirac spinor and vector representations are related through
formula~\rref{Sgam} in the main text
\[
{\cal S}^{-1} \gamma_{s}~{\cal S}= \Lambda_{s}^{~p} ~\gamma_{p}.
\label{spin}
\]

To prove that the Weyl spinor representations are integral
we use a theorem presented in~\cite{AS2} where it was shown that the whole 
group $SO(d,d\,|{\bf Z})$ is generated by a special 
subset of group elements. We will construct explicitly the Weyl spinor 
representation matrices  corresponding to the group elements in that  
subset and show that they are integral.
The subset contains three types of elements. The first type are  generators
of the form 
\beqn
\left(
\matcc
I_d & n \\
0   & I_d
\emat
\right),~~n^{T} =-n~~. \label{vectorone}
\eeqn
The second type of generators forming 
a $SL(d,{\bf Z})\times {\bf Z}_2$ subgroup
have the form
\beqn
\left(
\matcc
R & 0 \\
0   & {R^{T}}^{-1}
\emat
\right),~~\det R = \pm 1. \label{vectortwo}
\eeqn 
These are the T-duality generators corresponding  to a change of 
basis of the of the compactification lattice.

The final generator is given by
\beqn
\left(
\begin{array}{cccccc}
 0 & & &1& & \\
   &0& & &1& \\
   & &I_{d-2}& & &0_{d-2} \\
  1& & &0& & \\
 & 1& & &0& \\
  & &0_{d-2}& & &I_{d-2}
\end{array}
\right). \label{vectorthree}
\eeqn 
It corresponds to T-duality along the $x^1$\, and $x^2$\, 
coordinates.
The full duality group is in fact $O(d,d\,|{\bf Z})$ but here 
we only consider its restriction to  $SO(d,d\,|{\bf Z})$ which is
the subgroup that does not exchange Type IIA and IIB. The full T-duality 
group is then obtained by adding to the above  list one more generator
corresponding to  T-duality in a single direction.

Using~(\ref{Sgam}) 
one can check that the Dirac spinor representation corresponding to 
the first type of generator~(\ref{vectorone}) is
\beqn
\exp(\frac{1}{2} n_{ij} a_{i} a_{j}). \label{antn}
\eeqn
This has a finite expansion and is manifestly integer valued in the standard
Fock space basis obtained by acting with the creation operators on a 
vacuum state.

One can prove that the full $SL(d,{\bf Z})$ group is 
generated by its $SL(2,{\bf Z})_{ij}$ subgroups acting on
the $x^i$\, and $x^{j}$\, coordinates. We can use this to find 
the spinor representation matrices corresponding to generators of the second 
type~(\ref{vectortwo}). Since each 
$SL(2,{\bf Z})_{ij}$ is generated by its $T_{ij}$\ and  $S_{ij}$\, 
transformations,
which in the $(ij)$\, subspace where $i<j$ have the form 
\[
\left(
\begin{array}{cc}
1&1 \\
0&1 
\end{array}
\right),~~
\left(
\begin{array}{cc}
0&-1 \\
1&0 
\end{array}
\right),
\]
it is enough to
find the spinor matrices for these generators.
The spinor  representation of $T_{ij}$ is given by
\beqn
\exp(a_i a_j^{\dagger}). \label{spins}
\eeqn
The exponential~(\ref{spins}) 
has a finite expansion and its matrix elements are integer valued. 
Similarly the spinor  representation of $S_{ij}$  is given by
\beqn
\exp(\frac{\pi}{2}(a^{\dagger}_{j} a_{i}-a_i^{\dagger}a_j)).
\label{spint}
\eeqn
Let us define $A=a^{\dagger}_{j} a_{i}-a_i^{\dagger}a_j$ for fixed
values of $i$ and $j$. In terms of number 
operators $N_{i} = a_{i}^{\dagger} a_{i}$ we have 
 $A^2 = -N_i -N_j + 2N_i N_j$.
Since $N_i$ can be either zero or one, $A^2$ is zero or minus one. 
The Fock space can be split into a direct sum of
two subspaces, defined by the eigenvalues of $A^2$. On the subspace
defined by $A^2 = 0$, we also have $A=0$ and thus the spinor
representation~(\ref{spint}) reduces to the identity. On the subspace defined
by $A^2=-1$, the exponential can be written as 
$\cos(\pi/2)+A\sin(\pi/2)=A$. 
On both subspaces, the representation matrix of the transformation is
integer valued. 
A formula for the spinor representation of 
the $S_{ij}$ generators which is valid on both
subspaces is given by,
$1+A+A^2$. The second type of 
generator~(\ref{vectortwo}) 
also contains elements with $\det R=-1$. A spinor
transformation corresponding to such a generator 
is given by
\beqn
1-2a^{\dagger}_1 a_1. \label{detminus}
\eeqn

Finally, the generator~(\ref{vectorthree}) has the spinor representation
\beqn
\exp(\frac{\pi}{2}(a_1-a^{\dagger}_1)(a_2-a^{\dagger}_2)).
 \label{spintdual}
\eeqn
It  has a finite expansion given by 
$(a_1-a^{\dagger}_1)(a_2-a^{\dagger}_2)$, which can be obtained using
$((a_1-a^{\dagger}_1)(a_2-a^{\dagger}_2))^2=-1$, and in this form it 
is manifestly integral. 

Since the Fock space basis we have used splits into
two subsets of definite chirality, it follows that  the
Weyl spinor representations of $SO(d,d\,|{\bf Z})$ are also integral.

In the remainder of the appendix we show that the Weyl spinor 
representation of $SO(3,3|{\bf Z})$ is isomorphic to $SL(4,{\bf Z})$. 
First note that for the Lie algebra corresponding
to the continuous Lie groups we have the equivalence
$so(3,3\,|{\bf R})\cong sl(4,{\bf R})$.  The
spinor representation of the first group is isomorphic to the
fundamental of the second. Since in the first part of the appendix we proved
that the spinor representations are integral it is reasonable to expect that
they form a subgroup of $SL(4,{\bf Z})$. In fact we will show that they 
are isomorphic to the whole  $SL(4,{\bf Z})$ group.

We represent the Weyl spinor
state $n|0\rangle+\frac{1}{2}M^{ij}a^{\dagger}_{i}a^{\dagger}_{j}|0\rangle$
as the column
\beqn
\left(
\begin{array}{c}
n\\
M^{23}\\
M^{31}\\
M^{12}\\
\end{array}
\right).
\label{nM}
\eeqn
Using operators of the form~(\ref{spins}) and~(\ref{spint}) we 
generate an $SL(3,{\bf Z})$ subgroup of the form
\beqn
\left(
\begin{array}{cc}
1&0 \\
0&R
\end{array}
\right) \label{slthreez}
\eeqn
where $R$ is the same matrix appearing in~(\ref{vectortwo}). We 
will now show that the Weyl spinor representation also contains
$SL(2,{\bf Z})_{1i}$ subgroups which act on the first and the 
$i+1$ entries of the column spinor~(\ref{nM}). 
These subgroups together with~(\ref{slthreez}) generate the entire
$SL(4,{\bf Z})$ group. The T-duality 
generator~(\ref{spintdual}), denoted below $T_{12}$, 
has the Weyl spinor representation
\[
T_{12}=
\left(
\begin{array}{cccc}
0&0&0& -1 \\
0&0&-1&0 \\
0&1&0 &0 \\
1 &0&0&0
\end{array}
\right).
\]
Let us also consider a transformation $G$ given by
\[
G=
\left(
\begin{array}{cccc}
1&0&0& 0 \\
0& 1 &0&0 \\
0&0& a &b \\
0&0&c&d
\end{array}
\right),~~~~~ad-bc=1,
\]
which is an element of an $SL(2,{\bf Z})$ subgroup of elements of the
form~(\ref{slthreez}).
By conjugating $G$ with the $T_{12}$ generator
\beqn
T^{-1}_{12}GT_{12}=
\left(
\begin{array}{cccc}
d&c&0&0 \\
b&a&0&0 \\
0&0&1 &0 \\
0&0&0&1
\end{array}
\right), \label{dcba}
\eeqn
we find an $SL(2,{\bf Z})_{12}$ transformation acting on the first and
second entries. All the other  $SL(2,{\bf Z})_{1i}$ subgroups can be obtained
by conjugating~(\ref{dcba}) with elements of the form~(\ref{slthreez}).
Thus we have found Weyl spinor representations generating 
the entire $SL(4,{\bf Z})$ group. In fact the 
representation is isomorphic to $SL(4,{\bf Z})$ since  all the
spinor representation matrices~(\ref{antn}), (\ref{spins}), (\ref{spint}), 
(\ref{detminus}) and (\ref{spintdual}) are integral and have unit determinant.

\end{document}